# Megasonic Enhanced Electrodeposition

Jens Kaufmann[1], Marc P.Y. Desmulliez[1], Dennis Price[2]

[1] MicroSystems Engineering Centre (MISEC), School of Engineering & Physical Science,
Heriot Watt University, Edinburgh, EH14 4AS, United Kingdom

[2] Merlin Circuit Technology LTD, Harwarden Industrial Park Manor Lane,
Deeside, Flintshire, West Wales, CH5 3QZ, United Kingdom

*Abstract* A novel way of filling high aspect ratio vertical interconnection (microvias) is presented. High frequency acoustic streaming at megasonic frequencies enables the decrease of the Nernst-diffusion layer down to the sub-micron range, allowing thereby conformal electrodeposition in deep grooves. Higher throughput and better control over the deposition properties are therefore possible for the manufacturing of interconnections and metal-based MEMS.

## I. INTRODUCTION

The increasing consumer demand for faster, lighter and smarter electronic devices calls for enhanced system integration and packaging technologies. Key to the increasing density of electronic components is the introduction of the high density interconnection (HDI) technology in printed circuit boards (PCB), resulting in multilayer technology and increasing amounts of electrical connections that need to be handled. Microvias are elements that are particularly important as they allow the reduction of the footprint of electronic components through the redistribution of interconnects in the underlying layers. Microvias are formed by mechanical drilling or laser ablation of the PCB material and subsequent electroplating is used to fill the cavity to render them electrically conductive. The microvias need however a conductive seed layer on the side walls to allow the electrodeposition of metal. The technologies used in that respect are mainly autocatalytic plating processes or direct metallisation[1].

To allow further integration and miniaturization, it is necessary to form microvias with an aspect ratio (height over diameter hole ratio) greater than 1:1. This current limit arises because of the difficult hydrodynamic conditions and current crowding effect at the mouth of the hole; both conditions attenuate the convection in the electrolytic solution near the surface of the substrate. This lack of effective agitation reduces the ion concentration in the solution within the immediate proximity of the microvia, increases the Nernst diffusion layer and limits therefore the deposition rate of the metal [2]. To overcome this limitation, extensive research utilising a variety of additives and current waveforms was performed. The additives used are mainly large organic inhibitors, and small complex builders that accelerate the deposition at the bottom of the via cavity. This is a very complex method, which needs a carefully controlled solution. However, super conformal plating of deep trenches and holes can be achieved, as exampled by the damascene process metallisation. The use of this process in high volume consumer electronics would nevertheless be problematic because of the high parameter variants found in the usually large volumes of solution used in PCM manufacturing. This paper proposes therefore to use megasonic agitation to reduce the diffusion layer and achieve optimal electrodeposition for microvias with aspect ratio larger than 1:1.

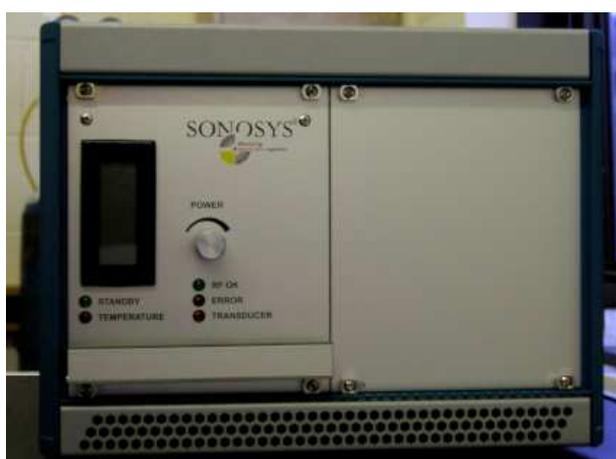

Fig. 1 Megasonic generator.
500 W Electrical power at 1MHz

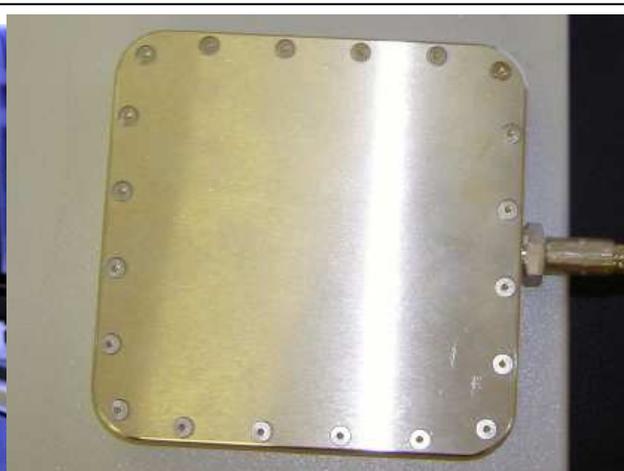

Fig. 2 Megasonic transducer of 4 by 4 inch active area

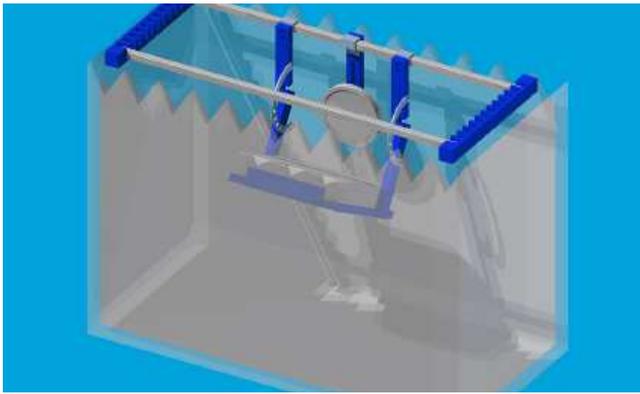

Fig. 3 Schematic of the transducer mounted on the carrier focussing the sound towards the sample holder

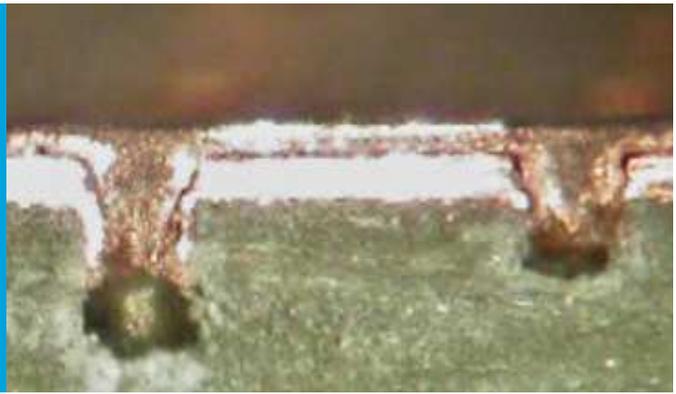

Fig. 4 Insufficient (left) or no filling (right) of the microvias

## II. MEGASONIC AGITATION

This paper describes a new method to enhance the capability and compatibility of metallisation processes based on megasonic assisted copper electrodeposition. The primary focus of this article concerns the filling of high aspect ratio microvias. However, this technology is transferable to the manufacture of various high aspect ratio metal MEMS structures as well as the development of photoresist in deep trenches. A high frequency acoustic streaming of frequency f allows the modification of the near surface hydrodynamics. The electrical potential of the ox-red reaction is governed by the activity of the ion near the electrode surface, according to the Nernst equation:

$$E = E^0 + \frac{RT}{z_e F} \ln \frac{a_{OZ}}{a_{Red}} \qquad (1)$$

Increasing the activity, $a$, and therefore the concentration, $c$, near the surface of the substrate, results in a decreased overpotential.

$$a_{M^{Z=}} = \gamma \cdot c(M^{Z+}) \qquad (2)$$

$$\frac{dc}{dx} = \frac{c_0}{\delta} \qquad (3)$$

More precisely, the Nernst diffusion layer which, in normal conditions, is governed by the velocity of the medium stream over the solid surface, as in (4), see [3], depends on $1/f^{1/2}$ in the presence of an acoustic field, as in (5).

$$\delta_{Hydrodynamic} = 0.16 \left( \frac{v}{Ux} \right)^{\frac{1}{2}} \qquad (4)$$

$$\delta_{acoustic} = \left( \frac{2v}{\omega} \right)^{\frac{1}{2}} \qquad (5)$$

The high frequency (above 500 KHz) not only reduces the thickness of the Nernst diffusion layer but avoids the creation of cavitations as traditionally experienced in ultrasonic agitation. These highly volatile collapsing bubbles inhibit the usage of acoustic streaming in the plating industry, due to the likelihood of board delamination. The use of acoustic streaming on the diffusion layer can be found in cleaning processes for micro parts. Furthermore the modelling of the megasonic induced flow in trench like cavities is also described in [4]

## II. EXPERIMENTAL WORK

In order to achieve the filling of microvias of high aspect ratio, a full Design of Experiments (DOE) has been conducted to analyse the influence of megasonic agitation on the plating quality of the microvias. The considered parameters include the current waveform, the power of the transducer used and the included angle of radiation between the transducer and the PCB. These parameters were chosen due to previous work in microvia plating. This abstract focuses at this stage on the fundamental differences introduced by using direct current (DC), pulsed current (PP) and reverse pulsed current (RP) plating. The different power levels for the megasonic agitation range between 0W and 500Watts of electrical power to distinguish between the normal agitation within the tank and the acoustic streaming. To change the angle between the transducer and the test vehicle (cathode) a custom carrier for the transducer was designed and manufactured. This carrier allows changing the radiation angle without changing the distance between the transducer and the centre of propagation.

The transducer itself consists of a 4 by 4 inch wide PZT crystal plate contained in a laser welded container. The experiments were conducted in a 60 litres polypropylene tank. The electrolyte contains the standard components for copper deposition; 10% sulphuric acid 300 g/l copper sulphate, 0.1% HCL and 0.05% of organic brighteners.

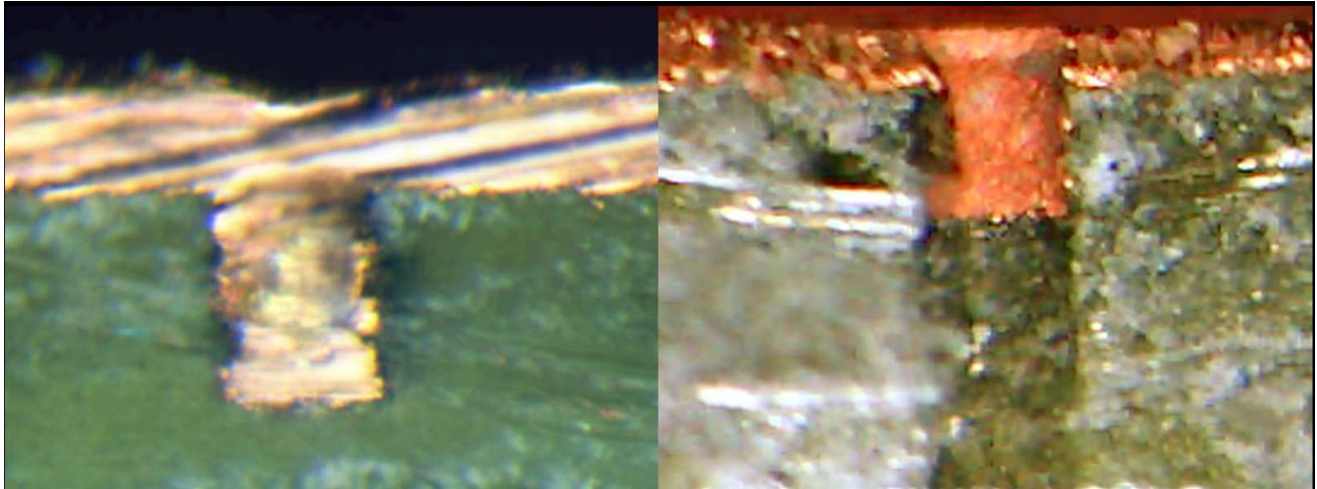

Fig. 5 Completely filled via (left) using megasonic agitation (125 W) with an aspect ratio of 2:1, uniformly plated vias with insufficient seed layer (right)

As measure of the quality of the microvias, several criteria were specified. The throwing power and the amount of deposited metal per unit time were measured. Finally, the highest achievable aspect ratio was recorded. The filling of the microvia was possible till an aspect ratio of >2:1 was reached. The filling profile of the holes shows a largely over proportional deposition at the bottom of the blind vias. Larger aspect ratio could not be achieved because of the lack of seed conductive layer deep into the microvia cavity. This result was obtained for 125 and 250 W of radiated power, at any angle and with a DC waveform. Pulse electroplating is not necessary.

### III. Conclusions

These results demonstrate the world's first application of megasonic agitation for the plating of microvias of aspect ratio greater than 1:1. The improvement in the deposition rate, as well as the quality of the deposits, make megasonic agitation a serious contender for the filling of deep aspect ratio microvias. The throwing power of the plating, the relation between the thicknesses on the mouth to the thickness on the bottom reached 5. This allows super conformal filling, as known from the damascene process in semiconductor manufacturing. This method does not require any special plating chemistry or optimization of waveform parameters. The effect of megasonic agitation on the crystalline structure needs however further investigation.